\documentclass[12pt,twoside]{article}   %For printing on two sides of page
\usepackage[super,sort,comma]{natbib}
\usepackage{fancyhdr}		%Gives headers and footers defined below
\usepackage{epstopdf}
\newcommand{\captionv}[3]{\begin{center}\parbox{#1cm}{\caption[#2]{{\sf #3}}}
        \end{center}}
\usepackage[section]{placeins}   %
\usepackage{graphicx}

\makeatletter \renewcommand\@biblabel[1]{$^{#1}$} \makeatother
\newcommand{\note}[1]{\mbox{}\\ \noindent \rule{16cm}{0.5mm} \\
{\em #1} \\ \noindent \rule{16cm}{0.5mm}
\typeout{    }
\typeout{***********note active on this page *************************}
\typeout{Note: #1  }
\typeout{****************************************end Note}
}

\newcommand{\cen}[1]{\begin{center} #1 \end{center}}

\usepackage{hyperref}
\hypersetup{ colorlinks,
    citecolor=blue,
    filecolor=blue,
    linkcolor=blue,
    urlcolor=blue
}

\usepackage{xcolor}
\definecolor{gray}{rgb}{0.6,0.6,0.6}
\definecolor{red}{rgb}{0.85,0,0}
\definecolor{green}{rgb}{0,0.85,0}
\definecolor{blue}{rgb}{0,0,0.85}
\definecolor{beige}{rgb}{0.92,0.87,0.78}
\usepackage[all]{hypcap}    %causes link to figures to go to figure, not caption
\usepackage{multirow}
\usepackage{textcomp}
\usepackage{tabularx}

\usepackage{booktabs}
\usepackage{amsmath,amssymb,amsfonts}

\usepackage{amsmath}
\usepackage{algorithmic}
\usepackage{graphicx}

\begin{document}

\cen{\sf {\Large {\bfseries Dual-energy Computed Tomography Imaging from Contrast-enhanced Single-energy Computed Tomography } \\
\vspace*{10mm}
Wei Zhao$^{1\dagger}$, Tianling Lyu$^{1,2\dagger}$, Yang Chen$^{2*}$, Lei Xing$^{1*}$} \\
$^1$Department of Radiation Oncology, Stanford University, Stanford, CA, USA\\
$^2$Department of Computer Science and Engineering, Southeast University, Nanjing, Jiangsu, China
\vspace{0mm}\\
%Version typeset \today\\
}
\pagenumbering{roman}
\setcounter{page}{1}
\pagestyle{plain}
$^{*}$Author to whom correspondence should be addressed. email: chenyang.list@seu.edu.cn; lei@stanford.edu \\
% note, probably best not to use a student's e-mail as it won't be valid for
% very long.

\begin{abstract}
\noindent {\bf Purpose:} In a standard computed tomography (CT) image, pixels having the same Hounsfield Units (HU) can correspond to different materials and it is therefore challenging to differentiate and quantify materials. Dual-energy CT (DECT) is desirable to differentiate multiple materials, but DECT scanners are not widely available as single-energy CT (SECT) scanners. Here we purpose a deep learning approach to perform DECT imaging by using standard SECT data.\\
{\bf Methods:} We designed a predenoising and difference learning mechanism to generate DECT images from SECT data. The method encompasses two parts: the first is image denoising using a fully convolutional network (FCN). The FCN is trained on the AAPM Low-Dose CT Grand Challenge dataset and applied to a set of contrast-enhanced routine DECT data to reduce image noise. The second is dual-energy difference learning using a U-Net type network. In this part, the denoised low-energy CT images together with the difference image between the low-energy image and its corresponding high-energy counterpart image, are used as the network input and output, respectively. Finally, the predicted difference image is added to the input low-energy image to generate noise correlated high-energy image. The performance of the deep learning-based DECT approach was studied using images from patients who received contrast-enhanced abdomen DECT scan with a popular DE application: virtual non-contrast (VNC) imaging and contrast quantification. Clinically relevant metrics were used for quantitative assessment. \\
{\bf Results:} The absolute HU difference between the predicted and original high-energy CT images are 1.3 HU, 1.6 HU, 1.8 HU and 1.3 HU (corresponding maximum absolute HU differences are 3.0 HU, 2.9 HU, 3.1 HU, and 3.0 HU)for the ROIs on aorta, liver, spine and stomach, respectively. The aorta iodine quantification difference between iodine maps obtained from the original and deep learning DECT images is smaller than 1.0\%, and the noise levels in the material images have been reduced by more than 7-folds for the latter. \\
{\bf Conclusions:} This study demonstrates that highly accurate DECT imaging with single low-energy data is achievable by using a deep learning approach. The proposed method allows us to obtain high-quality DECT images without paying the overhead of conventional hardware-based DECT solutions and thus leads to a new paradigm of spectral CT imaging. \\

\end{abstract}
\vspace{-2mm}
\note{$^{\dagger}$ W. Zhao and T. Lyu contributed equally to this work.}

\newpage     %may or may not be needed

%
%The table of contents is for drafting and refereeing purposes only. Note
%that all links to references, tables and figures can be clicked on and
%returned to calling point using cmd[ on a Mac using Preview or some
%equivalent on PCs (see View - go to on whatever reader).
\tableofcontents

\newpage

\setlength{\baselineskip}{0.7cm}      %double spacing		

\pagenumbering{arabic}
\setcounter{page}{1}
\pagestyle{fancy}
\section{Introduction}

\label{sec:introduction}
Computed tomography (CT) accounts for over 88 million clinical adult scans in the United States each year and represents one of the most important imaging modalities in modern medicine~\cite{al2020multiphase}. The modality, however, falls short in accurate quantification of tissue material composition because the pixel value or Hounsfield Units (HU) in a CT image represents the effective linear attenuation coefficient, which is an averaged contribution of all materials or chemical elements within the pixel and image pixels having the same CT number can correspond to different material compositions. To alleviate the problem, dual-energy CT (DECT) \cite{alvarez1976energy,mccollough2015dual}, which acquires two sets of CT data with different energy spectra, has been realized as a result of advancements in hardware and post-processing capabilities \cite{kalender1986evaluation,flohr2006first,johnson2007material,boll2008calcified,niu2014,petrongolo2018single,van2018image,shi2020}. DECT, which is also known as a major implementation of "spectral imaging", takes advantage of the energy dependence of the linear attenuation coefficients of the tissue to yield material-specific images, such as blood, iodine, or water maps. Several important clinical applications, such as virtual monoenergetic imaging \cite{yu2012dual,pomerantz2013virtual}, virtual noncontrast-enhanced images \cite{takahashi2008dual,ferda2009assessment,ho2012characterization,mangold2012virtual}, urinary stone characterization \cite{primak2007noninvasive,leng2015feasibility}, automated bone removal in CT angiography \cite{sommer2009value,schulz2012automatic}, perfused blood volume quantification \cite{johnson2006dual,meinel2014first}, and particle stopping power prediction~\cite{yang2010,hunemohr2013,shen2018,lapointe2018robust,lee2019,zhang2019experimental,landry2019,wohlfahrt2020} are made possible by the advent of DECT.

\begin{figure}[t]%[\sidecaptionrelwidth][t]
\centering
\includegraphics[width=\linewidth]{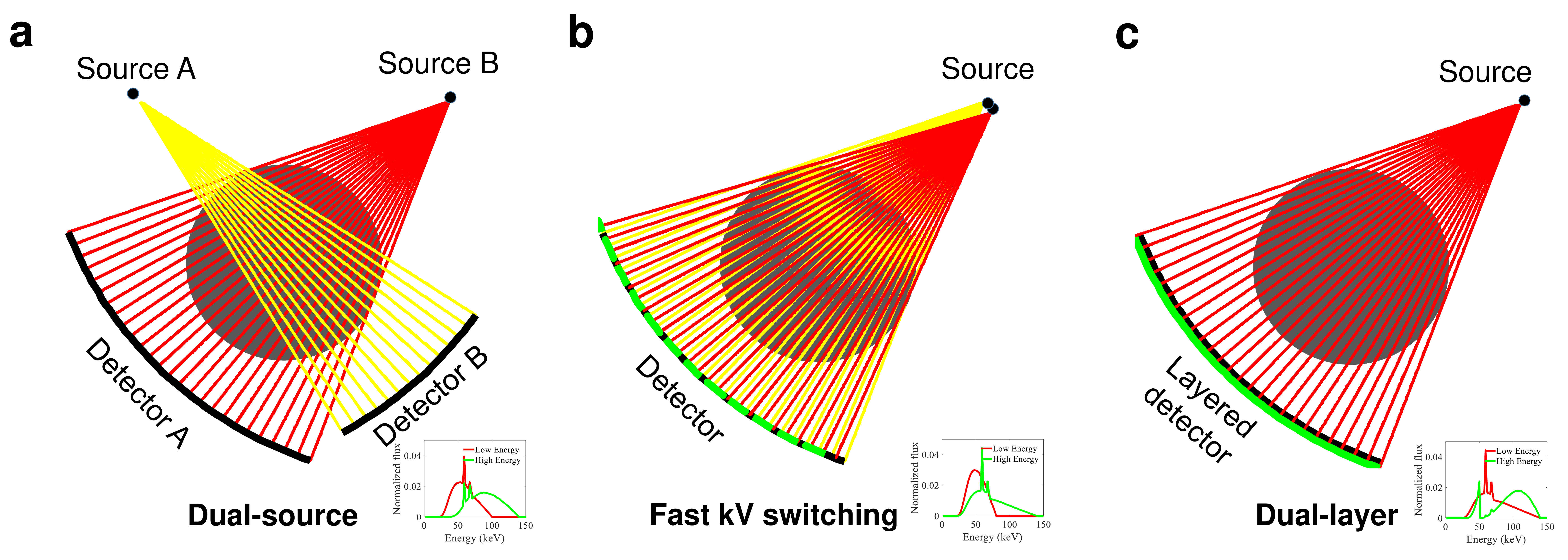}
\captionv{12}{}{A schematic diagram of different DECT imaging strategies. \textbf{a}, Dual-source DECT implemented by Siemens Healthineers. Two source-detector pairs are orthogonally mounted on the same gantry to acquire low- and high-energy projection data within a single CT scan. \textbf{b}, Fast kV switching DECT technique adopted by GE Healthcare, which acquires projections of low and high-energy spectra alternatively using one source-detector system. The tube potential of the X-ray source rapidly (within a millisecond) switches between the low- and high-kV settings. \textbf{c}, Dual-layer DECT approach adopted by Philips Healthcare scanner, which acquires DECT projection data using a layered detector, with the low-and high-energy data collected by the front and back-detector layer, respectively.\label{fig:1}}
\end{figure}

Different implementations have been realized for DECT imaging with different spectral information (Fig.~\ref{fig:1}), and all the implementations are highly challenging and represent milestones in the history of diagnostic CT and even in modern medicine. While these hardware-based solutions are capable of providing needed information for material composition analysis using either projection-domain or image-domain methods, the DECT system techniques are proprietary for the CT vendors and premium DECT scanners are usually more expensive than standard SECT scanner. Thus, DECT scanners are less accessible than SECT scanners, especially in less developed countries or regions. Deep neural network has attracted much attention for its unprecedented ability to learn complex relationships and incorporates existing knowledge into an inference model \cite{xing2018artificial,yang2018low,maier2019learning,shan2019competitive,shen2019patient,lee2019development}. Recently, it has been applied in DECT imaging, with successfully implementations in material decomposition \cite{liao2018pseudo,xu2018projection,zhang2019image,poirot2019physics} and virtual monochromatic imaging \cite{cong2017monochromatic,feng2018fully}. Previous preliminary studies also show DECT image with accurate CT value can be obtained from SECT image \cite{li2017pseudo,zhao2019dual}.
Based on the deep neural network, in this study, we propose a highly accurate noise reduction technique and demonstrate a predenoising and difference learning mechanism can provide DECT images with clinical favorable noise texture by using SECT data.
%Here we report a deep learning (DL)-based DECT (DL-DECT) solution, which uses a single SECT image as input and generates the image at one or multiple preselected energies with accurate HU and clinical desirable noise texture.
With the method, clinical DECT applications, such as iodine quantification, virtual noncontrast-enhanced imaging, can be performed readily using a conventional SECT scanner without the overhead associated with a high-end DECT, leading to a paradigm shift in spectral CT imaging.

\section{Material and methods}

\subsection{Overview of Study Design}
\begin{figure}[t]%[\sidecaptionrelwidth][t]
\centering
\includegraphics[width=0.8\linewidth]{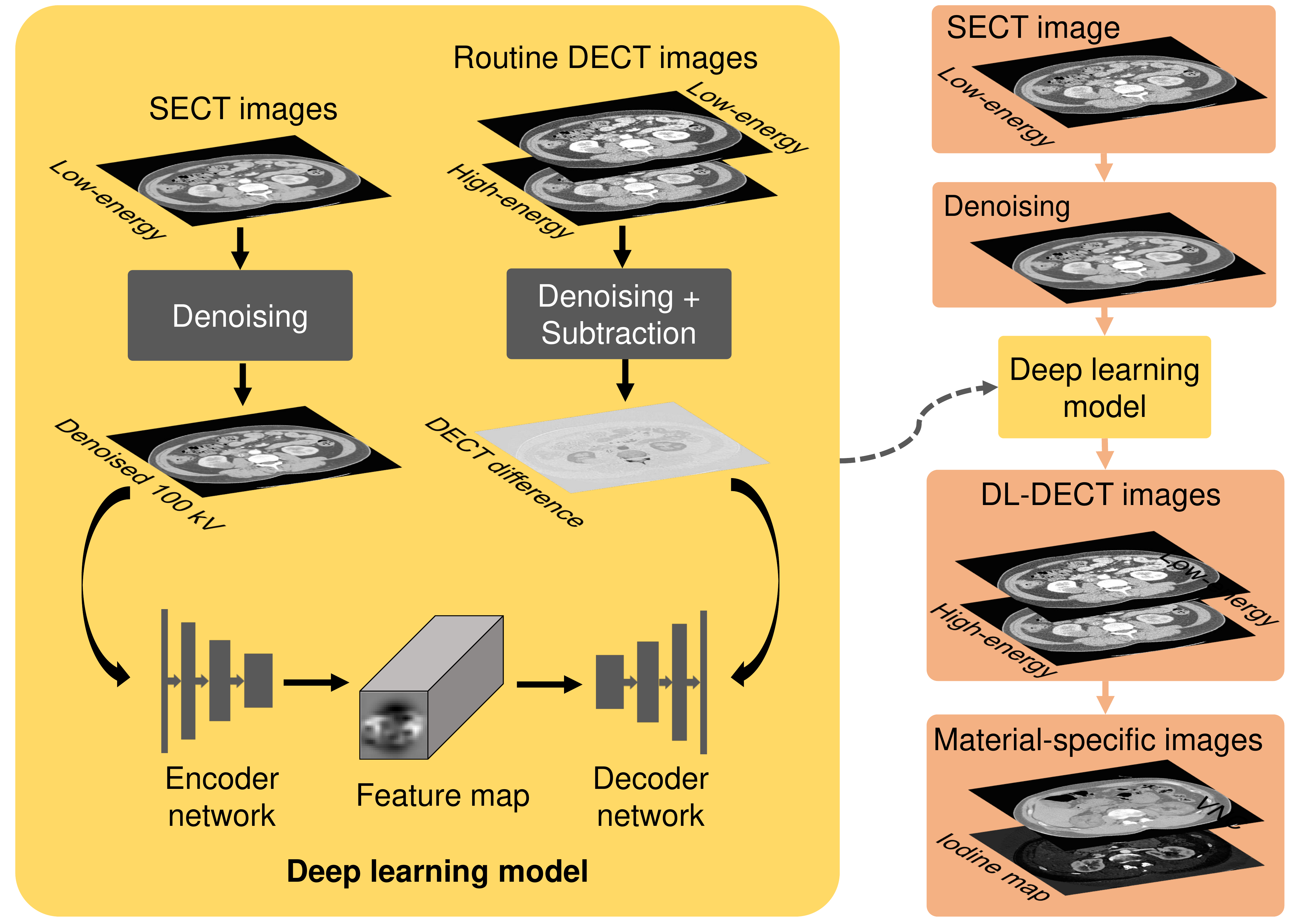}
\captionv{12}{}{Material decomposition using the proposed DL-DECT imaging. In the proposed data-driven strategy, denoised DECT images are fed into a deep neural network to learn the relationship between the dual-energy attenuation properties of the low- and high-energy CT images. With input single-energy CT images, the model is used to predict DECT images, which are subjected to material decomposition algorithm to obtain material-specific images.\label{fig:1c}}
%The training process learns a difference image between the low- and high-energy images. The model is implemented with a deep neural network architecture composed of an encoder network and a decoder network with skip connections. Its input is a denoised SECT image, and the output is its difference with respect to the corresponding high-energy image.
\end{figure}

The proposed deep learning (DL)-based DECT (DL-DECT) imaging strategy is shown in Fig. \ref{fig:1c}. The input to the model is the low-energy image $I_L$, whereas the output is, instead of the corresponding high-energy counterpart $I_H$, the difference between $I_L$ and $I_H$. By using the difference image $I_{diff}$= $I_L$-$I_H$ as the network output, the model extracts effectively the features characterizing the relationship between the training data pairs (i.e., the input low-energy images $I_L$  and the corresponding difference images $I_{diff}$) and integrate them into a predictive model (see~\ref{mapping} for details). The model is trained via supervised learning with routinely available DECT images. The mean-squared-error between the predicted and ground truth images is used as the loss function during the model training. To reveal the intrinsic energy-dependent attenuation properties of the tissues, both $I_L$ and $I_H$ are denoised by using a fully convolutional network (FCN) before being used for training and testing of the model (see~\ref{denoising} for details). With the trained model, a predicted high-energy CT image $I_H^{pred}$ can be obtained from SECT images of the patient by adding the predicted difference image $I_{diff}^{pred}$ to the raw low-energy image, i.e.,~$I_H^{pred}= I_{diff}^{pred}+I_L$.

The proposed approach is investigated by using 5,087 slices of clinical DECT images of 19 patients. All patients underwent routine contrast-enhanced abdominal DECT exams (100 kV/Sn 140 kV). To test the performance of the DL model, we use 960 slices of 100 kV images of 3 additional patients as input. The predicted 140 kV images $I_H^{pred}$ are assessed quantitatively by comparing the HU values with that of the ground truth images $I_H$ in region-of-interests (ROIs) of the testing abdominal CT scans. Virtual non-contrast (VNC) images and the iodine maps \cite{marin2014state} are derived to demonstrate the utility of the proposed method. Specifically, $I_H^{pred}$ together with the raw low-energy CT images $I_L$ are used to compute the VNC images and contrast quantification by using an image-domain material decomposition algorithm \cite{faby2015performance}. The material-specific images generated using the DL-DECT and the ground truth DECT are also compared using clinically relevant metrics. For the material-specific VNC images and iodine maps, the HU values, iodine concentrations and noise levels are compared across the body, especially in key organs including aorta, liver, spine and stomach. Statistical analysis were performed to compare DL predicted 140 kV images and raw 140 kV images.

\subsection{DECT Dataset}

In this retrospective study, clinical DECT images of 22 patients (19 males and 3 females; age [median]: 49 with range from 32 to 78;~age [mean$\pm$standard deviation]:~52.4$\pm$11.1) who had undergone iodine contrast-enhanced DECT exams between May 2013 and February 2016 were collected for the study. All the exams were performed in Nanjing General PLA Hospital, China, with the approval by the institutional review board. The DECT images were acquired using a SOMATOM Definition Flash DECT scanner (Siemens Healthineers, Forchheim, Germany) with $Abdomen\_DE$ scanning protocol after administering iodine contrast agent. The low- and high-energy of the DECT scans were 100 kV and 140 kV, respectively, and the 140 kV spectrum was pre-hardened using a tin filter for better spectral separation and dose efficiency. CT images were reconstructed using the filtered back-projection (FBP) algorithm (with the D30f convolutional kernel) provided by the commercial CT vendor. To train the DL-DECT imaging model, which yields the difference image between the low- and high-energy CT images for a given low-energy CT image, the DECT images of 16 randomly selected patients (which encompass 4736 image slices) were used. The DECT images of 3 patients (which encompass 1071 image slices) were used to tune the hyperparameters of the network during the validation phase, and the DECT images of the rest 3 patients were used to test the trained model. All slices were denoised before model training, validation and testing.

\begin{figure}[t]
\centering
\includegraphics[width=\linewidth]{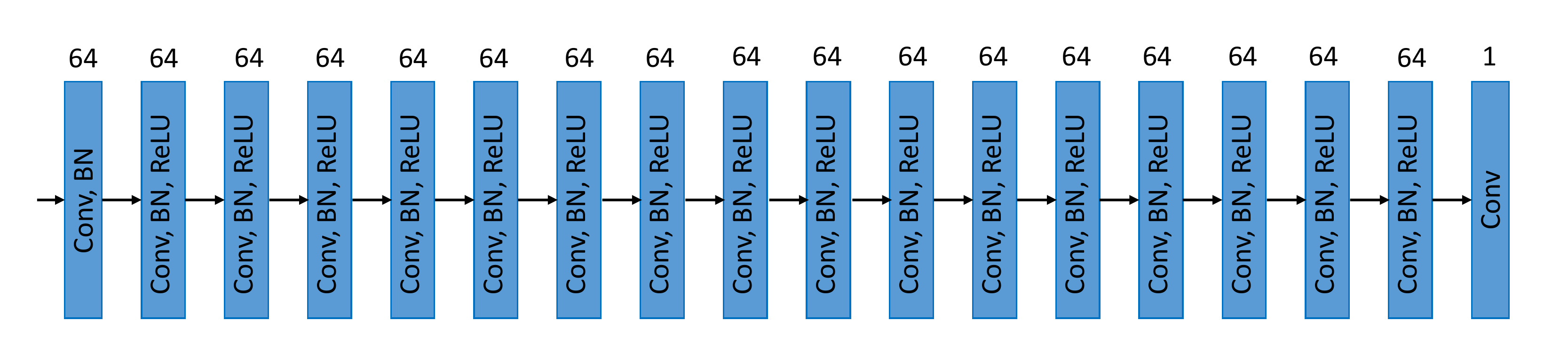}
\captionv{12}{}{Architecture of the FCN for image denoising. The input of the model is a standard DICOM CT image slice and the output is a noise reduced CT slice. The network encompasses an input layer, 16 convolution blocks and an output layer. Numbers atop each block show the number of channels of the multichannel feature maps. \label{fig:3}}
%The input layer performs 2D convolution operations followed by the batch normalization. Each of the 16 convolution blocks encompasses a set of convolution filters, followed by BN and a ReLU layer. The output layer performs convolution to generate the final activation map.
\end{figure}

\subsection{Denoising of DECT images}
\label{denoising}
DECT images are first denoised before used for training and testing to mitigate any adverse influence of noise. For this purpose, we developed and implemented an in-house FCN, which takes the raw CT image as the input and the difference image between the raw CT image and a low-noise (high-dose) CT image as the output. The neural network (Fig.~\ref{fig:3}) consists of an input layer, 16 convolution blocks and an output layer. The input layer (i.e., the first layer) conducts two-dimensional (2D) convolution operation using 64 kernels whose size is $3\times3$ and sliding stride is $1\times1$. Image padding is set to 1 such that the width and length of the convolutional feature map are the same as the input image. The size of the feature map generated using the input layer is $m\times n\times 64$ with $m$ and $n$ represent the width and length of the input image, respectively. Image patches $m=n=64$ are used for training, and standard DICOM CT images with $m=n=512$ are used for testing. The feature map is then fed into the 16 convolution blocks. Each of the convolution blocks encompasses a set of convolution filters, followed by batch normalization (BN) and a rectified linear unit activation (ReLU) layer. All the blocks have the same number of convolutional filters (64 filters), and all the filters have the same settings (kernel size $3\times3\times64$, stride $1\times1$, padding 1). By using this configuration, the sizes of the convolutional feature maps remain the same through the convolution blocks. The output layer uses a convolution kernel (size $3\times3\times64$, stride $1\times1$, padding 1) to generate the final activation map. With this network design, the output of the FCN has the same spatial shape as the difference image and can be compared with each other using a weighted $L2$ loss. The residual of the loss function is backpropagated to update the weights of the convolutional kernels of the FCN during the training procedure.

The denoising FCN model was trained using paired low- and high-dose CT dataset obtained from the American Association of Physicists in Medicine (AAPM) Low-Dose CT Grand Challenge (https://www.aapm.org/GrandChallenge/LowDoseCT/) which is hosted by Mayo Clinic. The dataset encompasses ten patient cases which were obtained on similar scanner models with use of automated exposure control and automated tube potential selection. Images reconstructed using projection data acquired using full dose are regarded as high-dose data. Poisson noise was inserted into the full dose projection data for each case to reach a noise level that corresponded to 25\% of the full dose. Images reconstructed using noise added projection are regarded as low-dose data. After the model is trained, the predicted difference image obtained during the inference procedure is subtracted from the raw CT image to yield images with significantly reduced noise.

The denoised DECT images are evaluated qualitatively and quantitatively using clinically relevant HU accuracy. Difference images between the raw and denoised images are calculated to show that whether anatomical information is well preserved after denoising. More importantly, for the contrast-enhanced abdominal CT examination, HU values of the raw and denoised images in some key organ tissues including the aorta, liver, spine and stomach are compared and quantitatively analyzed. To measure the HU values, a set of $11\times11$ pixels ROIs are selected in these organs. For each organ, the HU value is quantified using 5 ROIs at different locations.
%An example of the ROI selection is shown in Fig.~\ref{fig:2}\textbf{a}.(Fig.~\ref{fig:2})

\subsection{Predictive model for DL-DECT imaging}
\label{mapping}

%\begin{figure}[t]
%\centering
%\includegraphics[width=0.7\linewidth]{sf6.pdf}
%\caption{Examples of the input and output for the mapping network of dual-energy difference images. The input and output of the network are the denoised 100 kV CT images and the difference images between the denoised DECT images. All images are displayed with a window width (W=500 HU) and a center level (C=0 HU). }\label{fig:5}
%\end{figure}

\begin{figure}[t]
\centering
\includegraphics[width=\linewidth]{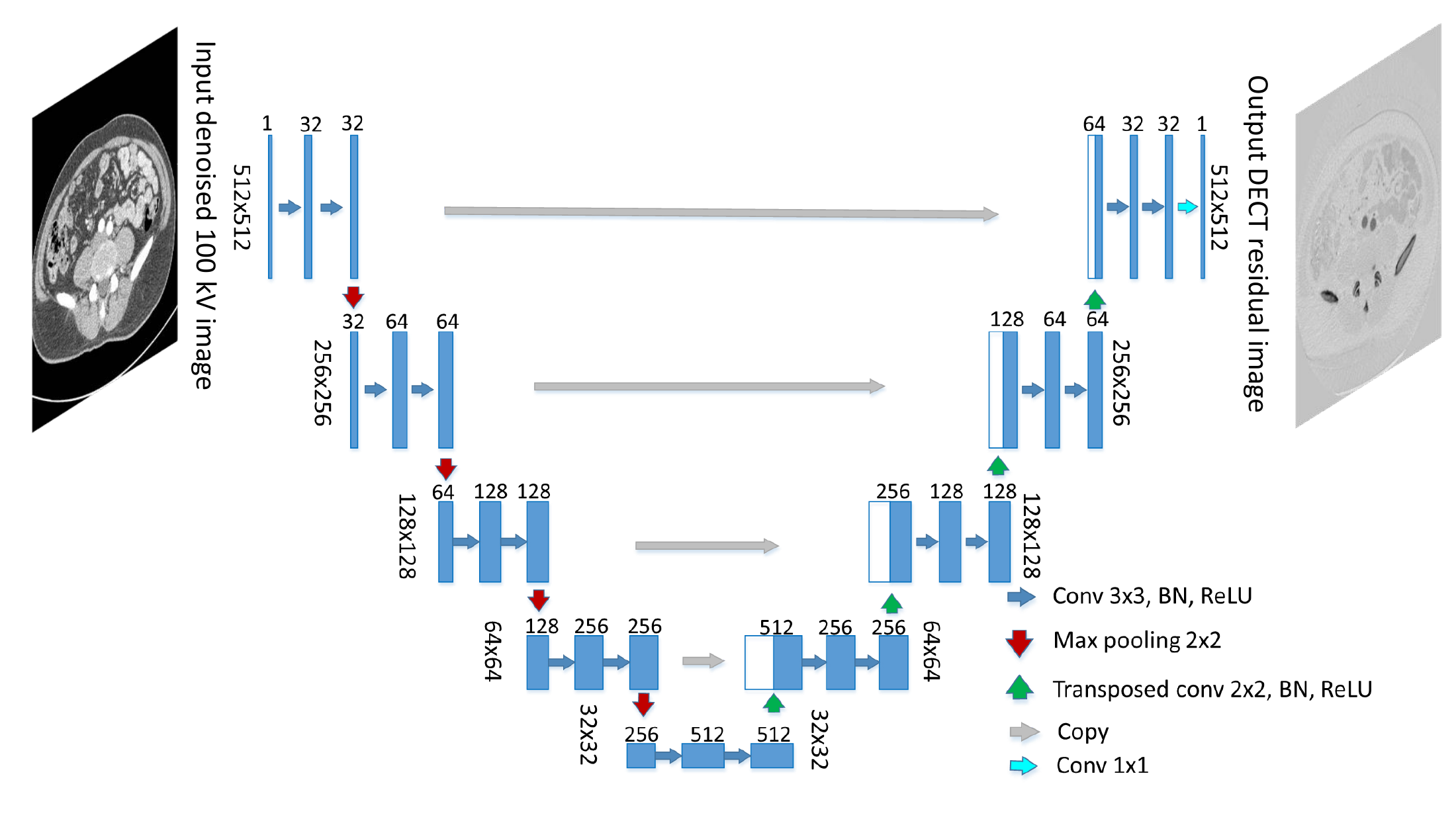}
\captionv{12}{}{The dual-energy mapping network for the DL-DECT imaging method. The U-net-type network uses an encoder-decoder architecture with skip connections and maps the denoised low-energy CT image to the difference image between the two DECT images. The input is a denoised low-energy CT image with $512\times512$ pixels. The output of the network is a difference image of the denoised DECT images of the same size as the input standard CT image. Numbers atop each network block are the number of feature channels. Numbers left to each block are spatial feature map shapes. \label{fig:4}}
\end{figure}

The denoised DECT images are used to train the DL mapping model for DL-DECT imaging. Instead directly yielding the high-energy CT image $I_H$ using the low-energy CT image $I_L$, we further construct an mapping model to take $I_L$ as the input and output the DECT difference image $I_{diff}$. The rationale for this design is that the network can be more focused on learning the difference between $I_L$ and $I_H$, resulting in more robust mapping. %Examples of the input and output images of the predictive model are shown in Fig.~\ref{fig:5}.and its architecture is delineated in Supplementary Fig. S1.
In detail, we use a U-net type network to infer $I_{diff}$. The network employs an encoder-decoder architecture with skip connections to find the difference mapping in an end-to-end fashion (Fig.~\ref{fig:4}). It has 26 hidden layers and an output layer. The hidden layers can be grouped into two parts. The first part (encoder) includes the first 14 hidden layers, and it takes a standard DICOM CT image as input to generate a multilevel, multiresolution feature representation by using successive convolution computation at multiple scales and abstraction levels.

The encoder incorporates five convolutional blocks and each of the blocks includes two consecutive 2D convolution layers. Each of the layers $conv(3,3,z,n)$ computes 2D convolutions using $n$ filters with a size of $3\times3$ for a given feature map with the number of channels $z$. The convolution layer is followed by a ReLU layer and a BN operation that normalizes the layer by adjusting and scaling the activations. By introducing the BN operation, we are able to fix the means and variances of each layer's inputs during the successive convolution computation. The convolutions are performed with a sliding stride of $1\times1$ and the same padding such that the spatial shape (height and width) of the feature maps does not change in the blocks. Each of the first 4 convolutional blocks is followed by a stride two max-pooling layer which performs down-sampling for the feature representation by a factor of two. The max-pooling layer is introduced to avoid over-fitting and to reduce the computational cost. Meanwhile, the number of the channels of the feature map for the five successive convolutional blocks has been increased from 32 to 512 by increasing the number of the convolutional filters by a factor of 2 for each block. Suppose the dimensions of the feature map is denoted as $m\times n\times k$ with $m$,$n$ and $k$ as the height, width and the channels of the feature map, respectively. Thus, for a given standard CT image, the dimensions of the feature maps change from $512\times 512\times 32$ (the first block) to $32\times 32\times 512$ (the fifth block) through the encoder.

The decoder consists of four convolutional blocks and four transposed convolutional operations. The convolutional block is the same as that in the encoder expect the number of the convolutional filters has been decreased from 512 to 32 by a factor of 2 for each block, resulting in the feature maps with a gradually reduced number of channels. Meanwhile, for each of the block, the input feature maps are concatenated correspondingly with the feature maps of the encoder that have the same resolution before going through the convolution layers (Fig.~\ref{fig:4}). The transposed convolution applies convolution with a fractional stride and results in an up-sampling by a factor of two. Hence, the height and width of the feature maps increased from 32 to 512, and the dimensions of the feature maps change from $32\times 32\times 512$ to $512\times 512\times 32$ through the decoder.

The output of the decoder is used as the input of the final layer, a $1\times1$ convolutional layer, to map the feature representation to the desired dual-energy difference. For training purpose, the output of the final layer is compared to the ground truth image (the corresponding difference image between the denoised 100 kV and 140 kV CT images) using the mean-squared error loss function.
This DL-DECT imaging network architecture enables a large reception field which is desirable for the dual-energy difference image mapping to preserve the high spatial resolution for DECT imaging.

%Figure \ref{fig:frog} shows an example of how to insert a column-wide figure. To insert a figure wider than one column, please use the \verb|\begin{figure*}...\end{figure*}| environment. Figures wider than one column should be sized to 11.4 cm or 17.8 cm wide. Use \verb|\begin{SCfigure*}...\end{SCfigure*}| for a wide figure with side legends.

\subsection{Network training}
Both image denoising network and the DL-DECT imaging network are trained on a workstation which consists of four Nvidia TITAN X GPUs (each with 12 GB GDDR5X memory) and 126 GB RAM. At the beginning of each training step, the network parameters in the convolution kernels are initialized randomly under Gaussian distribution with the mean value of zero and variance of $10^{-3}$. During the training phase, the loss function is minimized using the adaptive moment estimation (ADAM) optimizer and the parameters are updated using the back-propagation method.

The DL-DECT imaging network is implemented using Python 3.6 and Tensorflow deep learning library (version r1.9), together with cuda 9.0 and cuDNN 6 toolkit. The network is trained using 200 epochs and the learning rate is set to $10^{-3}$ for the first 150 epochs and $10^{-4}$ for the rest 50 epochs. The decay rates for the two learning rates in the ADAM optimizer are set to 0.9 and 0.999, respectively. A batch size of 4 is chosen to fully exploit the GPU memory. The training process takes about 24 hours for the difference mapping network.

\subsection{Material decomposition}

Dual-energy material decomposition exploits the energy-dependent information of DECT images and provides clinically important tissue composition information. In order to show the merit of the proposed DL-DECT imaging method, material decomposition is performed for the three testing cases and the results are compared quantitatively with that obtained using the standard DECT images. Two popular dual-energy applications, VNC imaging and iodine quantification, are also demonstrated using DL-DECT imaging. Briefly, a VNC image is an image in which the administrated contrast agent is artificially removed, mimicking a CT scan without contrast agent and possibly alleviating the need for CT scanning without contrast agent in clinical contrast-enhanced DECT imaging procedure. The iodine map (i.e., iodine concentration quantification) is commonly used to improve the subjective assessment of various lesions.

An image-domain material decomposition method widely used in clinical practice is applied to the DECT images to obtain the VNC images and iodine maps \cite{faby2015performance}. Using this method, the basis material images $\textbf{b}(\textbf{r})$ can be expressed as a weighted summation of the energy-resolved DECT images $\textbf{f}(\textbf{r})$. In a matrix form, it can be formulated as:
\begin{equation}\label{equ:1}
\textbf{b}(\textbf{r})=\textbf{W}\cdot \textbf{f}(\textbf{r}).
\end{equation}
Where $\textbf{W}$=\{$w_{me}$: \emph{m} the basis material and \emph{e} the energy level\} is the $2\times2$ weighting coefficients matrix. In order to obtain the material images, we need to calibrate the weighting coefficients matrix $\textbf{W}$. Since we aim to perform the material decomposition for a VNC image ($m=1$) and an iodine image ($m=2$), ROIs on water and iodine contrast agent of the DECT images are selected for calibration measurements. For the water ROI, it can be represented as $b_1=1$ and $b_2=0$, and the iodine contrast agent can be represented as $b_1=1$ and $b_2=1$. Hence, we have the following linear equations for the calibration measurements using the ROIs,
\begin{equation}\label{equ:2}
\textbf{b}_1=\textbf{W}\cdot \textbf{c}_1,~\rm{with}~\textbf{b}_1=\begin{pmatrix}
1 \\
0
\end{pmatrix}~\rm{and}~\textbf{c}_1=\begin{pmatrix}
c_{1L} \\
c_{1H}
\end{pmatrix},
\end{equation}
and,\\
\begin{equation}\label{equ:3}
\textbf{b}_2=\textbf{W}\cdot \textbf{c}_2,~\rm{with}~\textbf{b}_2=\begin{pmatrix}
1 \\
1
\end{pmatrix}~\rm{and}~\textbf{c}_1=\begin{pmatrix}
c_{2L} \\
c_{2H}
\end{pmatrix},
\end{equation}
\noindent here the $c_{1L}$  and $c_{1H}$ are the measured CT numbers of the water ROIs for the low- and high-energy CT images, respectively; $c_{2L}$  and $c_{2H}$ are the measured CT numbers of the iodine contrast agent ROIs for the two energies.  The weighting coefficients matrix $\textbf{W}$ can be calculated using Eqs.~\ref{equ:2} and \ref{equ:3} with the calibration measurements. Once $\textbf{W}$ is determined, material specific VNC image and iodine map can be reconstructed. Instead of using vendor provided software for material decomposition, the VNC images and iodine maps of the raw DECT and DL-DECT images were reconstructed using Eq.~\ref{equ:1} and the former one is used as a baseline comparison.

\subsection{Evaluation of the DL-DECT images}

To quantitatively evaluate the DL-DECT images and the consequent material decomposition images (VNC images and iodine maps), we compare the images obtained using the proposed approach with that obtained using the raw DECT images for three testing cases unseen during the model training process.

Clinically relevant metrics are used to assess the accuracy of the DL-DECT images. To measure the HU values, the ROIs with $11\times11$ pixels are selected in aorta, liver, spine and stomach. The HU values in a total of 60 ROIs in the DL-predicted and the raw 140 kV images are calculated and compared quantitatively. Difference images between the predicted and the raw 140 kV images in transverse, coronal and sagittal planes are also displayed to show the anatomical structure preservation of predicted images. Paired-sample $t$ tests were used for pairwise comparison between DL-predicted 140 kV images and raw 140 kV images. Statistical analysis was performed using Matlab 2017b software (MathWorks, Natick, Mass) with a statistically significant difference defined as $P<.05$.

Noise property of the DECT is important for material decomposition. To quantify the noise properties of the DECT images, in particular the correlation between the predicted and the raw 140 kV images in the context of the raw 100 kV images, we use the noise power spectrum (NPS) to quantify noise magnitudes at different spatial frequencies for the DECT images. The NPS is calculated from an ensemble average of the Fourier transform of noise-only images \cite{riederer1978noise,gang2012cascaded}. For the 2D CT slices,

\begin{equation}
%\begin{align*}
\rm{NPS}(f_x,f_y)=\frac{\Delta x \Delta y}{N_x N_y}\langle |\rm{DFT}\{ I(x,y)-\bar{I}(x,y)\}|^2\rangle,
%\end{align*}
\end{equation}

\noindent where $\Delta x=\Delta y=$ image pixel size, $N_x=N_y= 65$ (the width in pixels of each ROI for NPS measurement). The symbol $\langle\cdot\rangle$ denotes the ensemble average over all (n=10) ROIs, and DFT denotes discrete Fourier transform. The term $I(x,y)$ is the intensity of a ROI image under consideration and $\bar{I}(x,y)$ is the mean intensity value of the ROI image. The NPS are calculated for the DL-DECT and the raw DECT images.
To evaluate the accuracy of the VNC images and the iodine maps, we measure the HU value of the VNC image and the iodine concentration in the ROI inside the aorta.
%The HU values of the VNC images decomposed from the raw DECT and DL-DECT images agree each other well (Supplementary Fig. S9).  The iodine concentrations are also highly consistent, and more importantly, the noise levels of the iodine map obtained from the DL-DECT images are significantly lower than that obtained from the raw DECT images (Fig.~\ref{fig:4}).

\subsection{Comparison study}
The denoised DECT images are used in the training and testing of the DL-DECT imaging model. To demonstrate the benefit of the denoising procedure, a model trained by using raw DECT images (without noise reduction) is also established and the testing results are compared directly with that obtained after noise reduction. Specifically, the network is trained using the paired datasets of raw 100 kV images and the difference images between the raw 100 kV and 140 kV DECT images. The trained model is then deployed to the independent raw 100 kV testing images. The resulting 140 kV images are compared to the results obtained using the DL-DECT imaging model with denoised images. The difference images between the resulting 140 kV images and the ground truth are displayed and the HU values of the different types of organs are quantified.

\begin{figure}[t]%[\sidecaptionrelwidth][t]
\centering
%\vspace{-2mm}
\includegraphics[width=\linewidth]{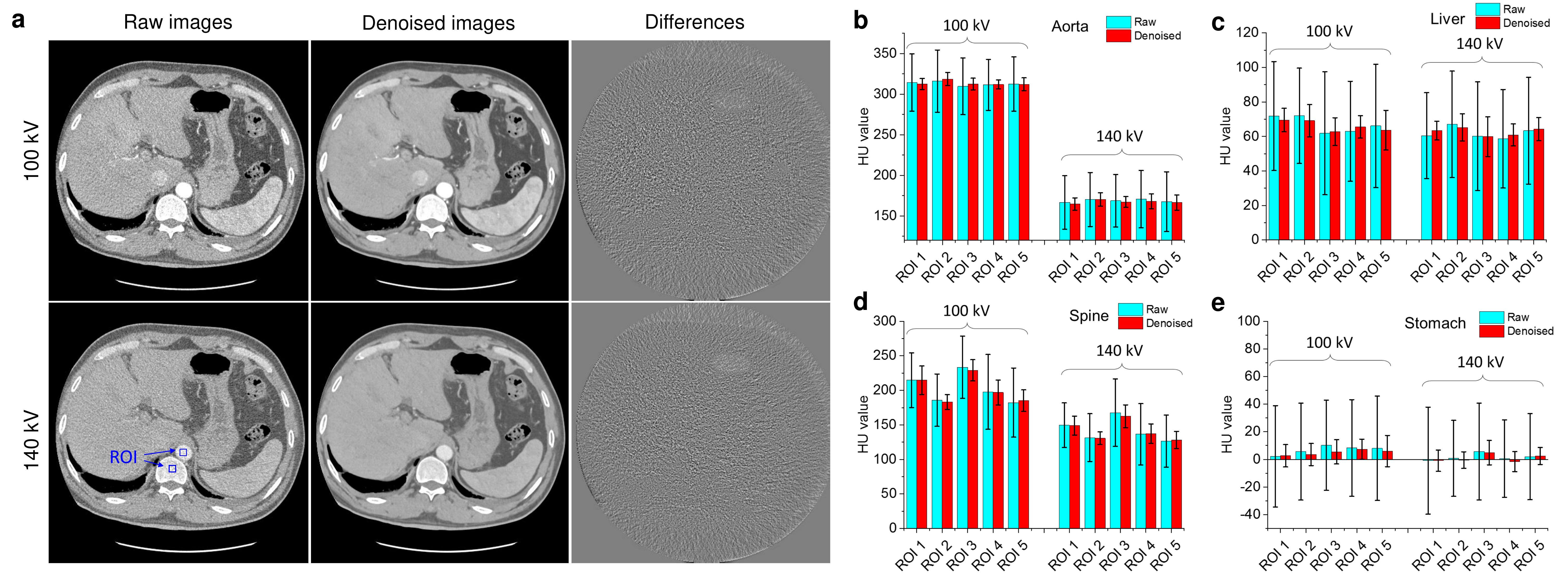}
\captionv{12}{}{Analysis of raw DECT images before and after FCN denoising for a testing patient. \textbf{a}, The 100 kV and 140 kV images before and after denoising. The difference between the images before and after denoising are also displayed for the two energies. No anatomical structure information is seen in the difference images, suggesting that the spatial resolution of the denoised CT images is well preserved. Two examples of ROIs used for quantitative assessment are depicted in the raw 140 kV image by blue boxes. \textbf{b-e}, Comparison of HUs of the DECT images in ROIs in the aorta, liver, spine and stomach before and after denoising. All CT images are displayed with a window width W = 500 HU and a center level C = 0 HU, while the difference images are displayed in a tighter window (C = 0 HU and W = 200 HU).\label{fig:5}}
%The denoised images are used for the training and testing of the DL-DECT imaging model.
\end{figure}

\section{Results}

\subsection{Denoising of DECT images}

For training and testing of the DL-DECT imaging model, the $I_L$ and $I_H$ are denoised by using the FCN technique first. The results in Fig. \ref{fig:5} show that the noise levels in the 100 kV and 140 kV DECT images are significantly reduced while the HU accuracy and anatomical details are well preserved. Fig.~\ref{fig:5}\textbf{a} shows the DECT images before and after noise reduction. The HU values of the denoised DECT images are consistent with the raw DECT images: the mean absolute HU difference between the two images is found to be 1.8 HU for all of the 60 ROIs for both low- and high-energy CT images (with a standard deviation of 1.3 HU). The noise levels are reduced by 3.9- and 3.7-fold for the 100 kV and 140 kV CT images, respectively (Fig. \ref{fig:5}\textbf{b-e}).

\begin{figure}[th!]%[\sidecaptionrelwidth][t]
\centering
\includegraphics[width=\textwidth]{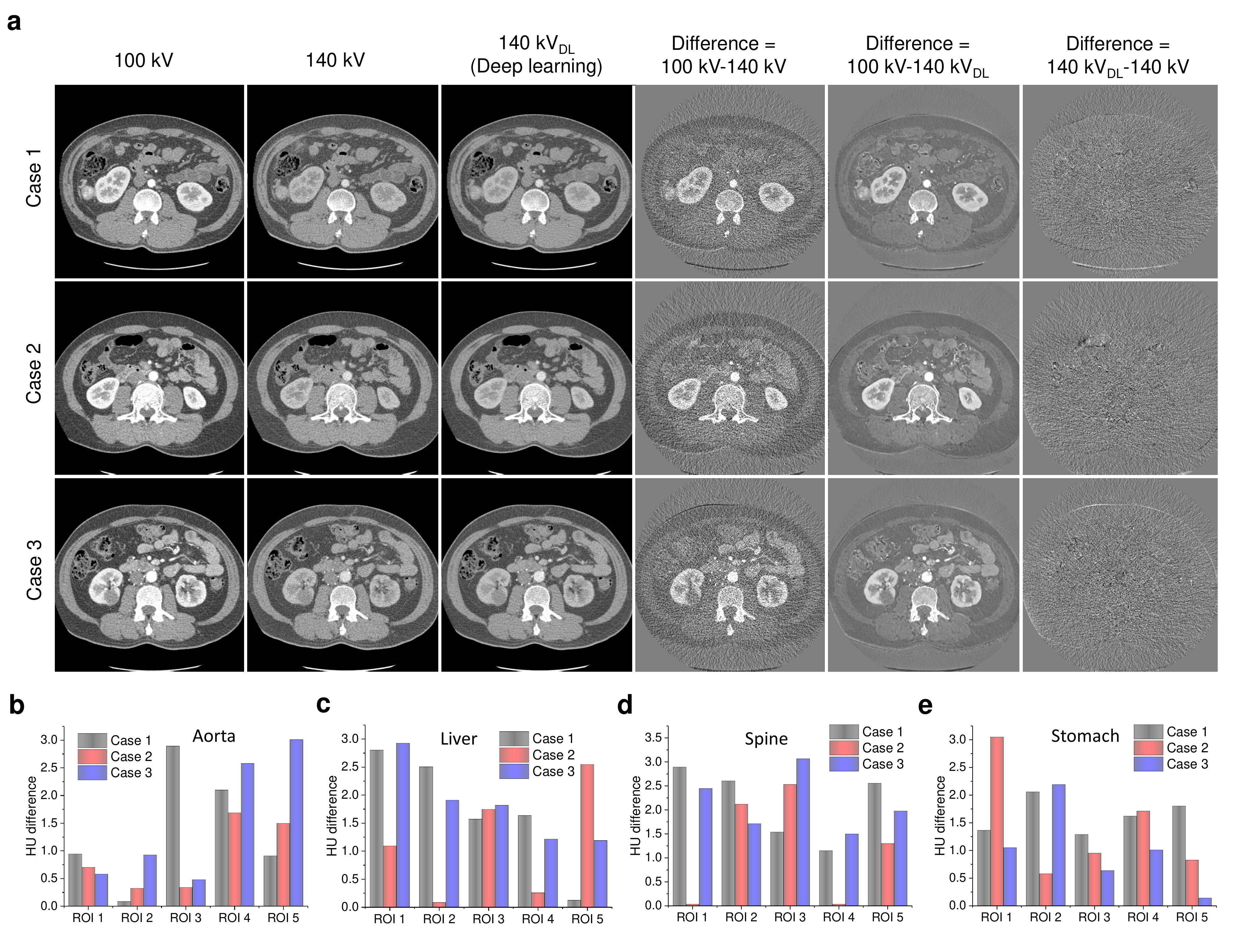}
\captionv{12}{}{Raw DECT and DL-DECT images. \textbf{a}, The first and second columns display the raw 100 and 140 kV CT images for three testing patients who underwent contrast-enhanced DECT imaging. The third and sixth columns show the DL-predicted 140 kV images and their differences with respect to the corresponding raw 140 kV images. For comparison, the fourth and fifth columns show the difference images between the raw DECT images, and the difference images between the raw 100 kV and DL-predicted 140 kV images. \textbf{b-e}, Absolute HU differences between the raw and DL-predicted 140 kV images for ROIs in aorta, liver, spine and stomach, respectively.  All CT images are displayed with a window width W = 500 HU and a center level C = 0 HU, while all difference images are displayed in a tighter window (C = 0 HU and W = 300 HU).\label{fig:6}}
%\vspace{-1em}
\end{figure}

\subsection{DL-DECT imaging}

The denoised DECT images are then used for the training and testing of DL-DECT. Fig.~\ref{fig:6} shows the raw DECT images (without denoising) and DL-DECT images for the three testing cases. Note that HU values of the iodine contrast in the 140 kV images are lower than that in the 100 kV images because the attenuation coefficient decreases with energy. In Fig.~\ref{fig:6}\textbf{a}, the first, second, and the third columns show the raw 100 kV, 140 kV, and the DL-predicted 140 kV images in transverse plane, respectively. The difference images between the DL-predicted and the raw 140 kV (ground truth) images are shown in the sixth column. For completeness, the difference images between the raw 100 kV and 140 kV images are shown in the fourth column along with the difference images between the raw 100 kV and the DL-predicted 140 kV images (the fifth column). It is seen that the DL-predicted 140 kV images are highly consistent with the ground truth images, with only some insignificant difference at the boundaries of anatomical structures, which may be due to the anatomical differences between the raw DECT images (the low- and high-energy data were acquired approximately $90^\circ$ out of phase using a dual-source DECT scanner). It is intriguing that the DL-DECT imaging model is capable of accurately generating the image content of the raw 140 kV images, even in the regions where the differences between the raw low- and high-energy DECT images are large (Fig.~\ref{fig:6}\textbf{a}, the 4th column). Figs.~\ref{fig:6}\textbf{b-e} show the absolute HU differences between the predicted 140 kV and the ground truth images in the ROIs in the aorta, liver, spine and stomach, respectively. For the aorta and spine, the five ROIs are from the upper to lower abdomen, whereas, for the liver and stomach, the five ROIs spread the organs. It is found that the HU values of the DL-predicted images in all ROIs agree with the ground truth values better than 3.1 HU, indicating the potential of the DL-DECT imaging model for clinical applications.

%\begin{figure}
%\centering
%\includegraphics[width=0.8\linewidth]{sf3.pdf}
%\caption{Quantitative evaluation between the raw and DL-predicted high-energy CT images. The absolute HU differences and standard deviations of the ROIs in aorta, liver, spine, and stomach are calculated placing the five ROIs on each of the three cases.}
%\end{figure}

\begin{table}%[b]%[tbhp]
\centering
\captionv{10}{}{Quantitative evaluation of the bone marrow and kidney regions with similar HU values in the low-energy CT images. For the bone marrow regions which have similar HU values as the contrast-enhanced kidney in the 100 kV images, the HU values of the bone marrow and kidney in DL-DECT 140 kV CT images show significant differences.}
\resizebox{\textwidth}{!}{\begin{tabular}{lrrr|rr|rr|rrr}
\toprule
\multicolumn{1}{ c }{\multirow{2}{*}{Unit: HU}} & \multicolumn{3}{ c }{100 kV (Raw)} &\multicolumn{2}{ c }{140 kV (Raw)} &\multicolumn{2}{ c }{140 kV (DL-DECT)} &\multicolumn{3}{ c }{P Value$^\dag$}\\ \cline{2-11}
 & KN$^\ast$ & BM$^\ast$ & P Value$^\dag$ & KN$^\ast$ & BM$^\ast$ & KN$^\ast$ & BM$^\ast$ & KN (DL vs Raw) & BM (DL vs Raw)& DL (KN vs BM)\\
\midrule
Case 1 & 204.4$\pm$11.5 & 205.9$\pm$10.9 & 0.28 & 117.9$\pm$28.4 &140.9$\pm$34.0 &	116.8$\pm$27.1 & 138.5$\pm$31.7 &0.76 &0.50& $<.00001$\\
Case 2 & 229.8$\pm$15.8 & 226.0$\pm$39.2 & 0.41 & 126.9$\pm$26.3 & 154.7$\pm$32.9 & 123.9$\pm$27.6 & 156.0$\pm$34.7 &0.70 &0.41&$<.00001$\\
Case 3 & 206.4$\pm$15.8 & 208.5$\pm$16.6 & 0.36 & 115.8$\pm$31.7 & 147.1$\pm$46.9 & 115.8$\pm$32.4 & 145.2$\pm$37.0 &0.69 &0.99&$<.00001$\\
\bottomrule
\vspace{0.1em}
\end{tabular}}
\label{tab:1}
\footnotesize{DL = Deep learning, DECT = dual-energy CT, BM = bone marrow, KN = kidney.\\
$^\ast$ Data are mean $\pm$ standard deviations.\\
$^\dag$ $P<.05$ is defined as the significance level.\\}
\end{table}
%\vspace{1em}

As shown in Table \ref{tab:1} for quantitative measurement and statistical analysis, although the contrast-enhanced kidney and some bone marrow regions have similar HU values in the 100 kV images, the HU values of the same bone marrow and kidney regions in the predicted 140 kV images have significant differences ($P<.00001$) which means the model has distinguished two different materials that have similar attenuation values at the low energy level and successfully outputs the DECT difference images, suggesting the nonlinear and spatial-dependent features of the DL-DECT imaging model. Note that the HU values of the bone marrow and kidney in DL-DECT 140 kV images show no significant difference with respect to that of the ground truth high-energy CT images. %(from 204 HU to 229 HU for the three testing cases)

\begin{figure}[th!]
\centering
\includegraphics[width=0.8\textwidth]{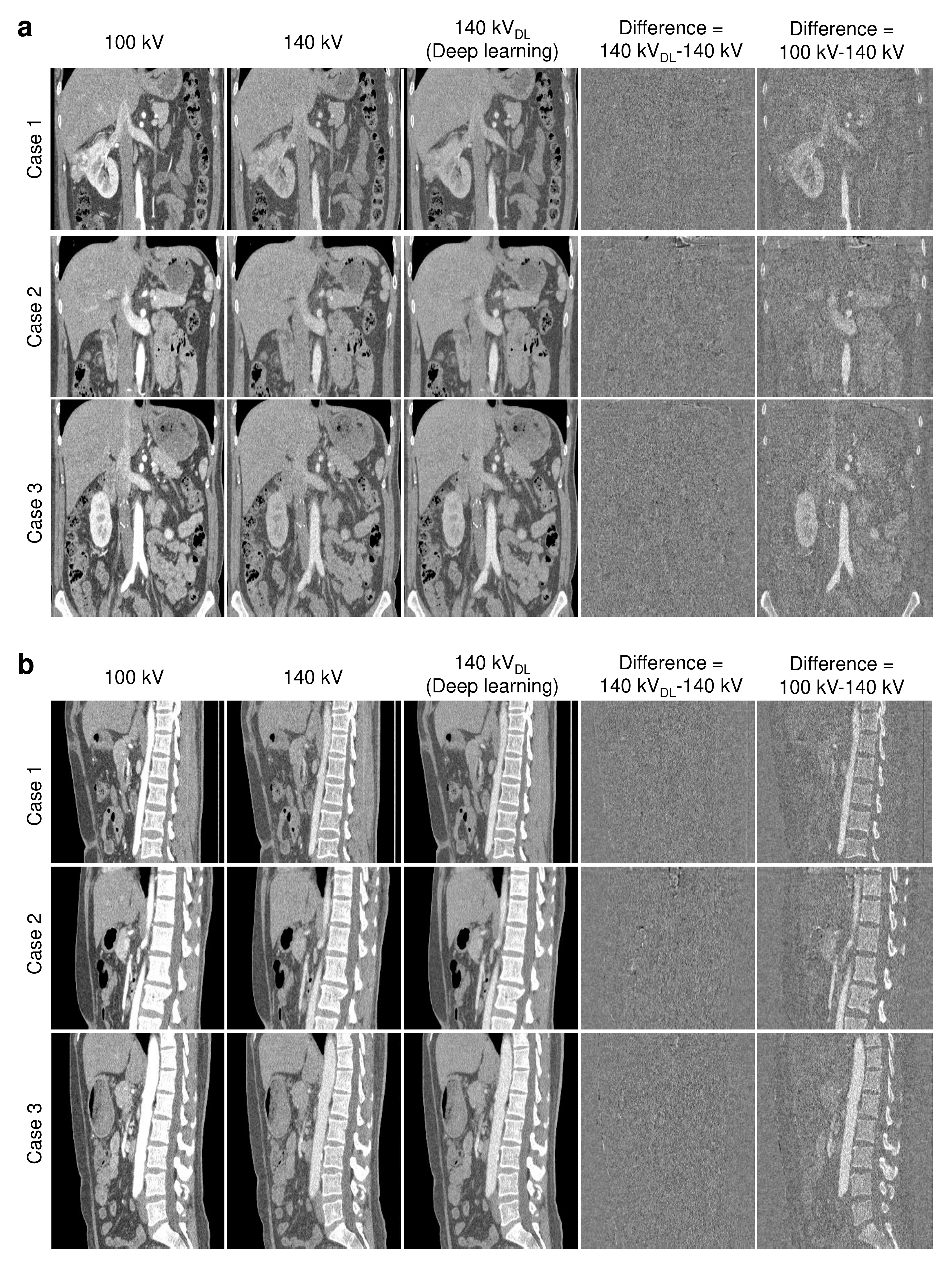}
\captionv{12}{}{Comparison between the raw DECT images and the DL-DECT images. a, Coronal plane; b, Sagittal plane. For both planes, the first and second columns show the raw 100 and 140 kV CT images of three testing patients, respectively. The third and fourth columns show DL-predicted 140 kV images and their differences with respect to the raw 140 kV images. All the CT images and the difference images are displayed with W=500 HU and C=0 HU. }\label{fig:coronal}
\end{figure}

Table~\ref{tab:2} (in Appendix) shows the comparison of the HU values in all ROIs in the DL-predicted and the raw 140 kV images. Statistical analysis indicates there are no significant differences between the DL-predicted images and the raw CT images. %The maximum absolute HU differences between the predicted and raw 140 kV images are 3.0 HU, 2.9 HU, 3.1 HU, and 3.0 HU for aorta, liver, spine and stomach, respectively.
The absolute HU differences between the DL-predicted and the ground truth 140 kV images are 1.3 HU, 1.6 HU, 1.8 HU, and 1.3 HU (corresponding maximum absolute HU differences are 3.0 HU, 2.9 HU, 3.1 HU, and 3.0 HU) for the ROIs in the aorta, liver, spine and stomach, respectively. Here the absolute HU difference for a given type of tissue is evaluated by averaging the values of all ROIs in the same organ of the 3 cases. The averaged absolute HU difference for each type of tissues is found to be less than 2 HU.
The comparison between the predicted and raw 140 kV images in coronal plane and sagittal plane are shown in Fig.~\ref{fig:coronal}.

\subsection{Material decomposition using DL-DECT images}

\begin{figure}[t]
\centering
\includegraphics[width=0.8\linewidth]{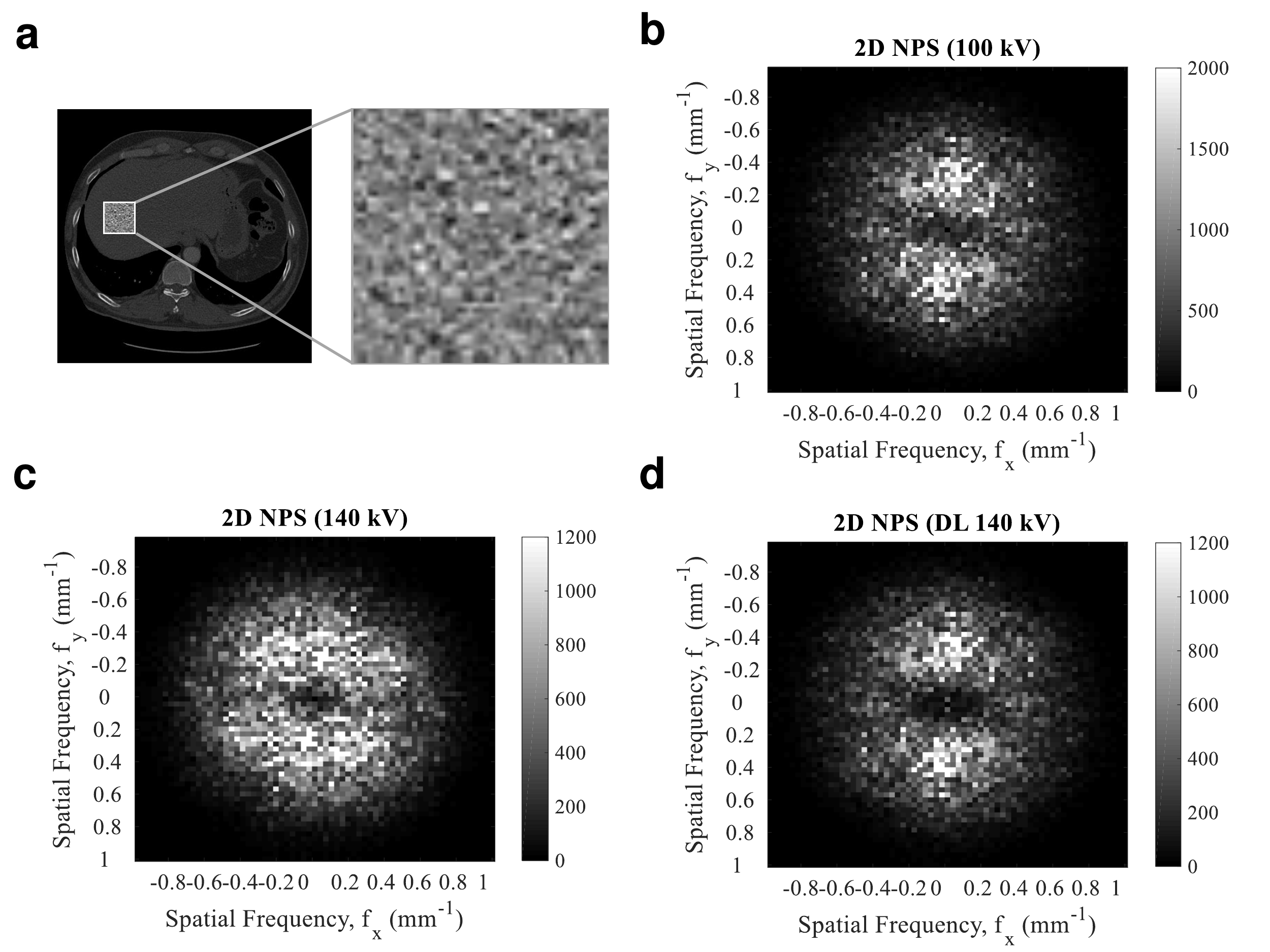}
\captionv{12}{}{Noise properties of the DL-predicted and raw DECT images. \textbf{a}, An example of the ROI used to generate noise-only images for NPS measurement. The ROI is $65\times65$ and selected on the uniform region of the liver. \textbf{b}, 2D NPS of the raw 100 kV CT image. \textbf{c}, 2D NPS of the raw 140 kV CT image. \textbf{d}, 2D NPS of the DL-predicted 140 kV CT image. \label{fig:8}}

%The NPS of the DL-predicted 140 kV image shows similar spatial structure as the NPS of the raw 100 kV image, but with a reduced magnitude. Since the 140 kV image is predicted using the 100 kV image, its noise texture is highly correlated with that of the raw 100 kV CT images. The NPS of the DL-predicted 140 kV image shows similar spatial structure as the NPS of the raw 100 kV image, but with a reduced magnitude. The noise correlation and the reduced magnitude are important in the DECT material decomposition, resulting in superior material-specific images.
\end{figure}

Fig.~\ref{fig:8} shows the NPS of the DL-predicted and raw DECT images. Since the 140 kV image is predicted using the 100 kV image, its noise texture is highly correlated with that of the raw 100 kV CT images. Compared to the raw 140 kV image, the DL-predicted 140 kV image shows similar spatial structure as the NPS of the raw 100 kV image and smaller NPS magnitudes at different spatial frequencies, suggesting material decomposition using the DL-DECT images can yields lower noise level.

%Since the 140 kV image is predicted using the 100 kV image, its noise texture is highly correlated with that of the raw 100 kV CT images. The NPS of the DL-predicted 140 kV image shows similar spatial structure as the NPS of the raw 100 kV image, but with a reduced magnitude. The noise correlation and the reduced magnitude are important in the DECT material decomposition.

\begin{figure}[th!]%[\sidecaptionrelwidth][t]
\centering
%\vspace{-1em}
\includegraphics[width=\textwidth]{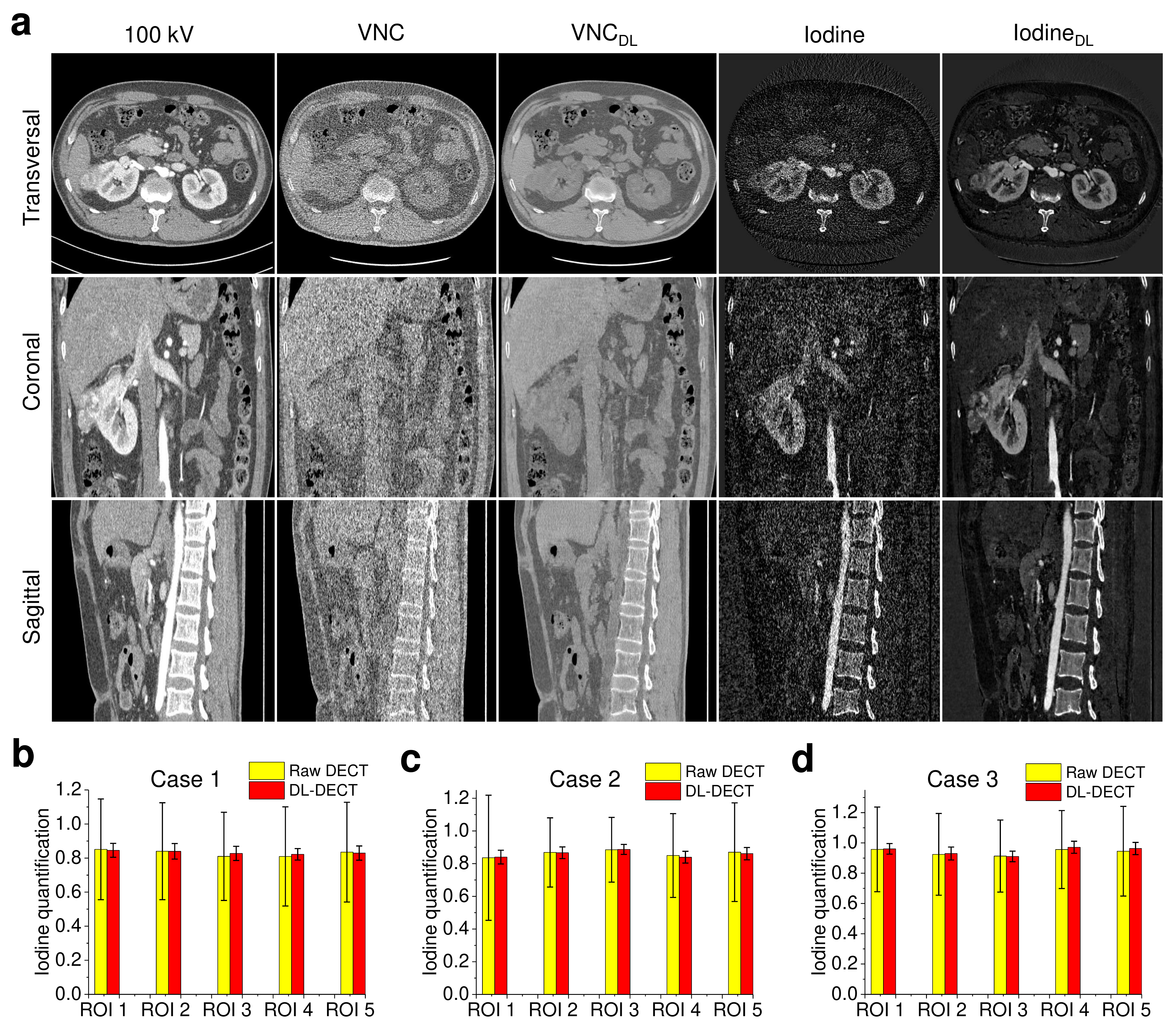}
%\vspace{-2mm}
\captionv{12}{}{Material decomposition using the raw DECT and DL-DECT images. \textbf{a}, Examples of the raw contrast-enhanced 100 kV CT images (1st column), VNC images (2nd column) and iodine maps (4th column) reconstructed using raw DECT images, VNC images (3rd column) and iodine maps (5th column) reconstructed using DL-DECT images in transversal, coronal and sagittal views. The 100 kV CT images and VNC images are displayed with a window width W = 500 HU and a center level C = 0 HU, and the iodine maps are displayed with W=1.2 and C=0.6. \textbf{b-d}, Quantitative comparisons of iodine concentration obtained using the conventional DECT and DL-DECT images. \label{fig:9}}
%\vspace{-2mm}
\end{figure}

Fig.\ref{fig:9} shows VNC images and iodine maps (Fig.~\ref{fig:9}\textbf{a}) as well as quantitative analysis results of material decomposition obtained by using the conventional DECT and DL-DECT images. For comparison, the raw 3D contrast-enhanced 100 kV CT images are also displayed. The VNC images in the second column and the iodine maps in the fourth column exhibit magnified noise levels as compared to the raw 100 kV CT images, primarily due to the adverse noise magnification effects of the matrix inversion during material decomposition. On the contrary, due to the correlation in the noise of DL-predicted 140 kV images and the raw 100 kV images, the DL-derived VNC images and iodine maps have remarkably reduced noise levels, showing superior image quality. %Compared to the VNC images and iodine maps derived from the raw DECT images, the results obtained based on DL show superior image quality and significantly reduced image noise.
The iodine concentrations in the aorta obtained using the raw DECT and the DL-DECT images agree each other based on the quantitative ROIs evaluation. Figs.~\ref{fig:9}\textbf{b-d} show the iodine concentrations of all ROIs in the aorta for the three contrast-enhanced testing cases. The iodine concentrations from the two types of images are consistent with each other, but the noise levels of the DL-derived iodine concentrations are significantly lower than that from the conventional DECT images.
The percentage differences of the iodine concentrations obtained using the raw DECT and DL-DECT images are 1.0\%, 0.6\%, and 0.9\% for the three testing cases, respectively. Meanwhile, the standard deviation in the ROIs, which characterizes the noise level, is reduced by more than 7-fold for all three cases.
%Figs.~\ref{fig:9}\textbf{e-g} show quantitative comparisons of the VNC images obtained using the conventional DECT and DL-DECT images. The HU values of the VNC images measured from the 5 ROIs placed in the aorta show consistent results.

The generation of the high-energy CT image is highly efficient once the DL-DECT imaging model is trained - it took about 5 seconds to map 300 CT slices (i.e., 16ms/slice) by using a standard PC platform (Intel Core i7-6700 K, RAM 32GB) with GPU (Nvidia GeForce GTX Titan X, Memory 12 GB).

\begin{figure}
\centering
\includegraphics[width=\textwidth]{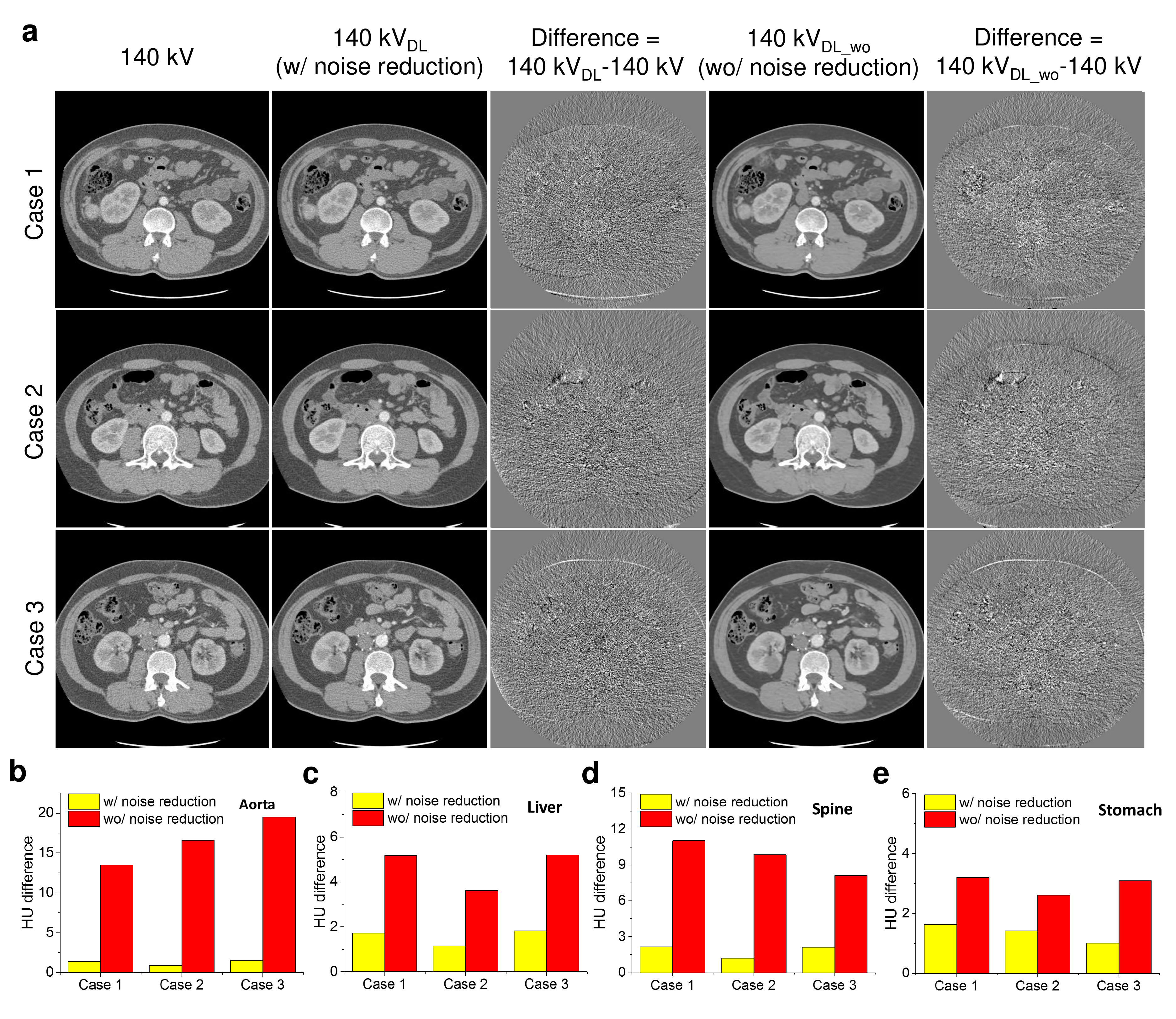}
\captionv{12}{}{The benefit of noise reduction in the mapping of dual-energy difference images. \textbf{a}, DL-predicted 140 kV CT images of three testing cases and their difference images with respect to the ground truths (1st column). The second and the fourth columns show DL-predicted 140 kV images with and without noise reduction, respectively. The third column and the fifth column show difference images between the predicted 140 kV images and the ground truths. All the 140 kV images are displayed with a window width (W=500 HU) and a center level (C=0 HU), while the difference images are displayed in a much tighter window (C=0 HU and W=200 HU) to better show the difference. \textbf{b-e}, Absolute HU differences between the raw and DL-predicted 140 kV images with and without noise reduction for ROIs in aorta, liver, spine and stomach, respectively.}\label{fig:noiseReduction}
\end{figure}

\begin{figure}
\centering
\includegraphics[width=\textwidth]{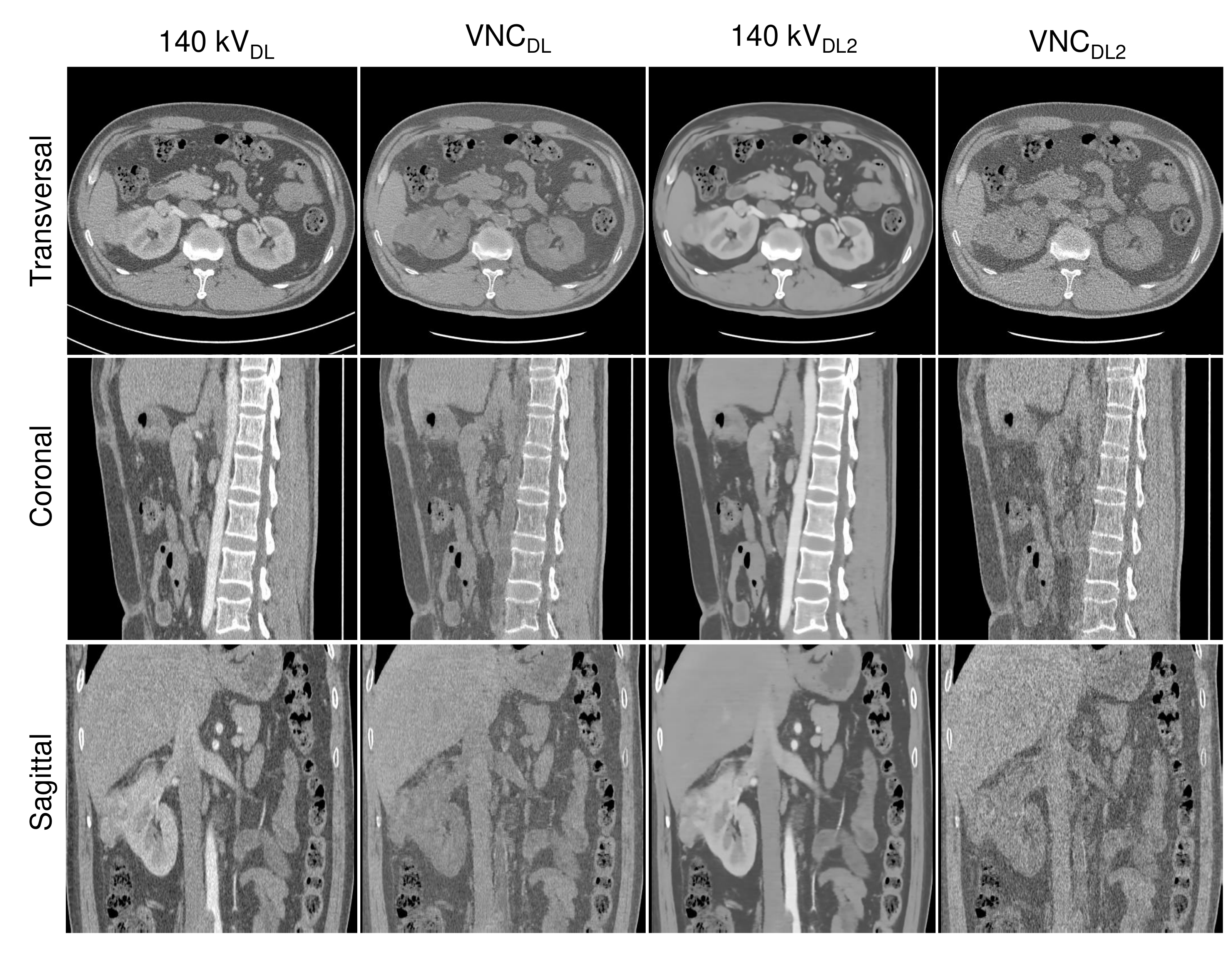}
\captionv{12}{}{Comparison of material decomposition using DL-predicted 140 kV images in two different options. The 140 kV$_{DL}$ images are obtained by adding the DL-predicted difference images to the raw 100 kV images, and the 140 kV$_{DL2}$ images are obtained by adding the predicted difference images to the denoised 100 kV images. VNC$_{DL}$ and VNC$_{DL2}$ are VNC images decomposed using 140 kV$_{DL}$ images and 140 kV$_{DL2}$ images, respectively. As can be seen, 140 kV$_{DL}$ images preserve noise texture. Although 140 kV$_{DL2}$ images show significantly reduced noise level, VNC$_{DL2}$ images show higher noise levels than VNC$_{DL}$ images. All the CT images are displayed with W=500 HU and C=0 HU. }\label{fig:12}
\end{figure}

\subsection{Comparison study}
Fig.~\ref{fig:noiseReduction} shows the quantitative comparisons between the performance of the DL-DECT imaging model trained and tested using DECT images with and without noise reduction. We find that the denoising procedure significantly enhances the accuracy of the predicted 140 kV images, whilst prediction without denoising provides inferior results.
The denoising procedure reduces the HU error for aorta, liver, spine, and stomach from 16.5, 4.7, 9.7, and 3.0 HUs to 1.3, 1.5, 1.8, and 1.3 HUs for the three testing cases, respectively. %were compared and quantitatively evaluated to show the

In the proposed predenoising and difference learning approach, the predicted difference image was added to the raw low-energy image (without denoising) to generate the final high-energy image. In order to further show the merit of the approach, we also performed study by adding the difference image to the denoised low-energy image. As shown in Fig.~\ref{fig:12}, in this case, the final DL predicted 140 kV images (denoted as $140~kV_{DL2}$) show a lower noise level than the 140 kV images obtained by adding to the raw low-energy images (denoted as $140~kV_{DL}$). However, the VNC images decomposed using the $140~kV_{DL2}$ images show a higher noise level than the VNC images decomposed using the $140~kV_{DL}$ images. This suggests the noises of $140~kV_{DL}$ images and the low-energy images have been reconciled during material decomposition due to their correlation, showing the merit of the approach.
%With the pre-processing of the images, the HU errors (averaged over the three testing cases) of aorta, liver, spine, and stomach are reduced from 16.5, 4.7, 9.7, and 3.0 HUs to 1.3, 1.5, 1.8, and 1.3 HUs, respectively (Supplementary Fig. S6), indicating the merit of the denoising procedure.

\clearpage
\section{Discussion}
The proposed DL-DECT imaging model learns the relationship between the low- and high-energy images and provides high-quality DECT images by using SECT data and the quality of the material decomposition maps derived from the DL-DECT imaging approach is remarkable.% and unseen before, even with the state-of-the-art DECT scanners from leading vendors.
%In reality, the proposed DL strategy can be implemented in a variety of ways. For example, the learning process from low-energy to predict high-energy images can be reverted by inputting the high-energy CT data to derive the low-energy CT image.
We note that all of our calculations are done in the image domain, which makes the implementation of the approach easy. NPS measurements suggest that the DL-predicted high-energy CT images preserve the original image noise texture and reduce noise magnitude. The noise texture of an image is important in diagnostic imaging as it critically affects the detection and characterization of subtle anatomical structures, especially in the low image contrast regions. Since human eye may unconsciously filter high-frequency image noise, it is desirable to preserve the original image noise texture.

The data-driven approach can learn complex relationships from training data (i.e., routine DECT images) and incorporates prior knowledge into an inference model. Hence, the model predicts the high-energy image not solely from the low-energy image, but also from the prior knowledge. The physics rationale of the approach can be explained as follows: for DECT images, the low- and high-energy images are spatially consistent and also correlated in energy-domain. For example, for CT scans using the same scanning protocol, the relationships of the HU values of the iodine contrast with different concentrations between the low- and high-energy images are fixed. The model can exploit the spatial consistency and energy-domain correction to learn the relationships of different materials between DECT images, especially the correlation of the iodine contrast agent for the contrast-enhanced CT scan. This is clearly demonstrated by the contrast-enhanced kidney region in Fig.~\ref{fig:6}. As can be seen, the iodine concentrations (gradually increase from the inner part to the outer part) at the kidney regions vary a lot. For all of the different concentrations in the kidney, the model has successfully predicted the difference image from the given 100 kV image, even for the pixels having similar HU values as the bone marrow regions in the 100 kV image (their HU values are significantly different in the 140 kV images (Table~\ref{tab:1})).

Two salient features of our approach are worth emphasizing here. First, the input CT images are denoised before sending to the DL-DECT imaging network for training and testing. By doing so, the adverse influence of the image noise is mitigated when constructing the predictive model and the intrinsic attenuation properties of the materials at high- and low-energy levels are learnt with high-fidelity, improving the robustness of the model. During the course of our study, we also tried to build a DL predictive model by using images without denoising and found that the resultant model yields much inferior DECT images (Fig. ~\ref{fig:noiseReduction}). Since residual noise presented in the denoised DECT images after denoising, the noise of the final 140 kV images is determined by both the predicted difference images and the raw 100 kV images, and should not be exactly the same as the latter. Meanwhile, the reduce noise magnitude in the predicted 140 kV images suggests the noise of the predicted difference images has been reconciled with the noise of the raw 100 kV images. Second, instead of directly predicting the high-energy CT images from the input low-energy data, our model is trained to predict the difference between the low- and high-energy images. Since CT images are always calibrated using the water images at the corresponding scanning protocols, water always has the same HU value in different energy level images (i.e., 0 HU). We found that the difference images characterize more accurately the change of tissue attenuation properties in the DECT images (such as bone and iodine contrast), which enables the mapping network to be focused on the region where there has significant HU difference. With this design, the model not only yields high-energy CT images with highly accuracy, but also preserves the noise texture of the input raw CT images which is favorable for DECT material decomposition (Fig.~\ref{fig:12}).

%This property also facilitates subsequent material decomposition.
%The noise preservation property is favorable for DECT material decomposition. In this predenoising and difference learning approach, the predicted difference image was added to the raw low-energy image (without denoising) to generate the final high-energy image. Another option is that the difference image can be added to the denoised low-energy image. As seen in Supplementary Fig. S7, by adding the difference images to the denoised 100 kV images, the final DL predicted 140 kV images (denoted as $140~kV_{DL2}$) show a lower noise level than the 140 kV images obtained by adding to the raw low-energy images (denoted as $140~kV_{DL}$). However, the VNC images decomposed using the $140~kV_{DL2}$ images show a higher noise level than the VNC images decomposed using the $140~kV_{DL}$ images. This suggests the noises of $140~kV_{DL}$ images and the low-energy images have been reconciled during material decomposition due to their correlation, showing the merit of the proposed method.

Due to beam hardening effect, the HU value of a specific pixel depends on the attenuation coefficient of this specific pixel, as well as the pixel location. The model for DL-DECT imaging performs prediction not solely based on the HU value of the input low-energy CT image, but also based on the pixel location (or spatial information). Pixels having similar HU values (contrast-enhanced kidney and some bone marrow regions) in the low-energy images were identified with their spatial information and then successfully mapped to their high-energy differences with accurate HU value accordingly, as shown in Fig.~\ref{fig:6}\textbf{a} and Table ~\ref{tab:1}. This also suggests the dual-energy mapping is not linear and a simple linear function or piece-wise linear function can not accurately map the low energy image to its corresponding high-energy by cosmetically changing the CT values.
In this study, we have used an in-house developed FCN method to provide noise significantly reduced DECT images. However, it is not mandatory to use the FCN method and other DL-based \cite{wolterink2017generative,kang2018deep,yang2018low,shan2019competitive} or classical image denoising methods can be directly applied to the proposed framework to replace the current FCN method. However, since different denoising methods have different performances (such as whether the fine-structures can be well preserved or not), the impact of these different denoising techniques on the performance of the DL-DECT needs comprehensive evaluations, which will be explored in the near future.

The proposed DL-DECT method starts with SECT data and yield highly accurate and noised correlated DECT images. These DECT images can be applied to standard DECT application, including material characterization, virtual monochromatic imaging, VNC imaging and so on. More importantly, conventional dual-energy material decomposition which usually uses matrix inversion yields amplified image noise. However, due to the noise correlation, the DL-DECT images can provide noise significantly reduced material-specific images and virtual monochromatic images, showing the merit of the proposed method. %Although we have used the DL approach to successfully extract some important features and characteristics from SECT images and provided accurate low- and high-energy CT images that typically obtained only using DECT scanner, by the same token, applying DL to DECT should potentially likewise reveal much additional information than the standard premium DECT scanners. %Also, we have used contrast-enhanced abdomen CT scan data to evaluate the proposed method. Of note, based on its nonlinear and location-dependent mapping features, it should be possible to apply the method to distinguish whether a hyperdensity is iodinated contrast agent or hemorrhage, which is important DECT application since DECT has been shown to be nearly 100\% sensitively and specificity for such discrimination.% \cite{gupta2010evaluation}.

Potential limitations of the DL-DECT imaging method should be discussed. First, current DL model is trained and tested using DECT images with the 100 kV/Sn 140 kV dual-energy scanning protocol. For a different dual-energy scanning protocol, one may need to train a new model. Moreover, the model depends implicitly on the beam spectra, making the mapping solution vendor-specific. On the other hand, one may argue that vendor-independent solution can be trained using CT images with carefully standardized calibration measurements or using an assemble DECT images from different vendors. Finally, the material decomposition is performed in image domain. As thus, artifacts caused by effects such as beam hardening and scatter may reduce the accuracy of the material-specific images of the proposed DL-DECT imaging method. Of course, similar to conventional material decomposition schemes in image domain which also suffer from the artifacts, various correction methods can be applied here to further enhance the accuracy of material decomposition.

%\vspace{-0.5em}
\section{Conclusion}

This study demonstrates that highly accurate DECT imaging and material-specific maps are achievable by using only SECT data via DL. Compared to the standard DECT techniques, the proposed method significantly simplifies the system design and reduce the noise level of the final material decomposition. The DL strategy allows us to obtain high-quality DECT images without paying the overhead of conventional hardware-based DECT solutions and thus lead to a new paradigm of spectral CT imaging. It may, in the future, find widespread clinical applications, including cardiac imaging, angiography, perfusion imaging and urinary stone characterization and so on.

% following only if there is an appendix
\newpage
\section*{Appendix}
\label{append}
\addcontentsline{toc}{section}{\numberline{}Appendix}

\begin{table}[ht!]\centering
\captionv{10}{}{Quantitative assessment of the DL-predicted and raw high-energy 140 kV images. Clinically relevant metric HU values are measured
using ROIs on aorta, liver, spine and stomach for all three testing cases. Statistical analysis indicates there are no significant differences
between the DL-predicted images and the raw CT images. The maximum absolute HU differences between the predicted and raw 140 kV
images are 3.0 HU, 2.9 HU, 3.1 HU, and 3.0 HU for aorta, liver, spine and stomach, respectively.}
\resizebox{\textwidth}{!}{\begin{tabular}{lcccc||lcccc}
\toprule
ROIs$^\ddag$ & Raw 140 kV$^\ast$  & DL 140 kV$^\ast$ & Error & P Value$^\dag$ & ROIs$^\ddag$ & Raw 140 kV$^\ast$ & DL 140 kV$^\ast$ & Error & P Value$^\dag$ \\
\midrule
Case1-Arota-ROI1 & 166.72$\pm$33.04 & 167.66$\pm$28.88 & -0.94 & 0.82 & Case2-Spine-ROI1 & 215.00$\pm$43.21 & 214.97$\pm$32.83 & 0.03 & 0.99 \\
Case1-Arota-ROI2 & 170.21$\pm$33.31 & 170.29$\pm$30.70 & -0.08 & 0.98 & Case2-Spine-ROI2 & 207.41$\pm$35.98 & 205.30$\pm$34.50 & 2.12 & 0.60 \\
Case1-Arota-ROI3 & 168.88$\pm$32.36 & 165.98$\pm$28.18 &  2.89 & 0.42 & Case2-Spine-ROI3 & 206.40$\pm$48.81 & 208.93$\pm$43.29 & -2.53 & 0.51 \\
Case1-Arota-ROI4 & 170.79$\pm$35.28 & 168.69$\pm$26.09 &  2.10 & 0.61 & Case2-Spine-ROI4 & 195.05$\pm$48.50 & 195.02$\pm$34.69 & 0.03 & 0.99 \\
Case1-Arota-ROI5 & 167.65$\pm$36.86 & 168.56$\pm$26.81 & -0.91 & 0.82 & Case2-Spine-ROI5 & 204.71$\pm$51.95 & 203.41$\pm$44.66 & 1.30 & 0.71 \\
Case1-Liver-ROI1 &  60.49$\pm$24.85 &  63.29$\pm$25.25 & -2.80 & 0.41 & Case2-Stomach-ROI1 & -0.21$\pm$30.74 &  2.83$\pm$18.67 & -3.05 & 0.41 \\
Case1-Liver-ROI2 &  67.05$\pm$30.90 &  64.55$\pm$20.03 &  2.50 & 0.43 & Case2-Stomach-ROI2 &  1.26$\pm$27.41 &  1.84$\pm$22.00 & -0.58 & 0.84 \\
Case1-Liver-ROI3 &  60.14$\pm$31.51 &  58.57$\pm$28.28 &  1.57 & 0.67 & Case2-Stomach-ROI3 &  0.89$\pm$28.28 &  1.84$\pm$18.95 & -0.95 & 0.75 \\
Case1-Liver-ROI4 &  58.69$\pm$28.53 &  57.05$\pm$23.13 &  1.64 & 0.63 & Case2-Stomach-ROI4 &  1.87$\pm$25.65 &  3.58$\pm$27.01 & -1.71 & 0.59 \\
Case1-Liver-ROI5 &  63.34$\pm$30.90 &  63.21$\pm$26.11 &  0.12 & 0.97 & Case2-Stomach-ROI5 & -1.95$\pm$27.36 & -1.12$\pm$23.79 & -0.83 & 0.82 \\
Case1-Spine-ROI1 & 149.50$\pm$32.21 & 152.40$\pm$30.06 & -2.89 & 0.35 & Case3-Arota-ROI1 & 189.83$\pm$31.85 & 189.26$\pm$24.19 & 0.58 & 0.88 \\
Case1-Spine-ROI2 & 131.51$\pm$34.56 & 134.12$\pm$31.13 & -2.60 & 0.43 & Case3-Arota-ROI2 & 185.43$\pm$31.51 & 184.50$\pm$24.33 & 0.93 & 0.80 \\
Case1-Spine-ROI3 & 167.66$\pm$48.86 & 166.12$\pm$34.73 &  1.54 & 0.76 & Case3-Arota-ROI3 & 187.85$\pm$29.91 & 188.33$\pm$25.72 & -0.48 & 0.89 \\
Case1-Spine-ROI4 & 136.43$\pm$44.36 & 137.58$\pm$43.61 & -1.15 & 0.80 & Case3-Arota-ROI4 & 198.64$\pm$32.81 & 196.06$\pm$29.07 & 2.58 & 0.47 \\
Case1-Spine-ROI5 & 126.42$\pm$37.77 & 128.98$\pm$39.11 & -2.55 & 0.52 & Case3-Arota-ROI5 & 186.83$\pm$34.33 & 183.83$\pm$30.73 & 3.01 & 0.47 \\
Case1-Stomach-ROI1 & -0.92$\pm$38.79 & 0.45$\pm$29.00 & -1.36 & 0.77 & Case3-Liver-ROI1 & 68.42$\pm$23.87 & 65.50$\pm$22.53 & 2.93 & 0.24 \\
Case1-Stomach-ROI2 &  0.80$\pm$27.49 & 2.86$\pm$27.51 & -2.06 & 0.53 & Case3-Liver-ROI2 & 59.37$\pm$32.38 & 57.46$\pm$24.93 & 1.91 & 0.61 \\
Case1-Stomach-ROI3 &  5.73$\pm$35.04 & 4.44$\pm$25.19 &  1.29 & 0.76 & Case3-Liver-ROI3 & 63.69$\pm$25.10 & 65.50$\pm$26.32 & -1.82 & 0.52 \\
Case1-Stomach-ROI4 &  0.55$\pm$28.02 & 2.17$\pm$27.88 & -1.62 & 0.67 & Case3-Liver-ROI4 & 62.02$\pm$30.52 & 63.23$\pm$26.27 & -1.21 & 0.71 \\
Case1-Stomach-ROI5 &  2.01$\pm$31.16 & 3.81$\pm$27.71 & -1.80 & 0.65 & Case3-Liver-ROI5 & 60.99$\pm$36.29 & 62.18$\pm$30.75 & -1.19 & 0.78 \\
Case2-Arota-ROI1 & 194.20$\pm$38.84 & 193.50$\pm$32.29 &  0.70 & 0.89 & Case3-Spine-ROI1 & 158.96$\pm$34.95 & 156.51$\pm$31.87 & 2.45 & 0.56 \\
Case2-Arota-ROI2 & 183.56$\pm$25.97 & 183.88$\pm$26.25 & -0.32 & 0.91 & Case3-Spine-ROI2 & 170.80$\pm$37.39 & 172.51$\pm$34.59 & -1.71 & 0.69 \\
Case2-Arota-ROI3 & 183.07$\pm$23.94 & 182.74$\pm$22.32 &  0.34 & 0.90 & Case3-Spine-ROI3 & 173.60$\pm$46.64 & 176.66$\pm$36.99 & -3.07 & 0.51 \\
Case2-Arota-ROI4 & 184.45$\pm$29.68 & 186.14$\pm$27.57 & -1.69 & 0.63 & Case3-Spine-ROI4 & 178.43$\pm$33.98 & 179.93$\pm$33.44 & -1.50 & 0.72 \\
Case2-Arota-ROI5 & 183.93$\pm$30.33 & 185.42$\pm$32.73 & -1.50 & 0.72 & Case3-Spine-ROI5 & 148.45$\pm$25.96 & 150.42$\pm$35.65 & -1.98 & 0.54 \\
Case2-Liver-ROI1 &  68.84$\pm$28.31 &  67.75$\pm$28.49 &  1.09 & 0.77 & Case3-Stomach-ROI1 & 6.43$\pm$31.60 & 7.48$\pm$24.71 & -1.05 & 0.74 \\
Case2-Liver-ROI2 &  63.36$\pm$34.43 &  63.27$\pm$24.48 &  0.08 & 0.98 & Case3-Stomach-ROI2 & 4.90$\pm$34.97 & 7.09$\pm$27.03 & -2.19 & 0.54 \\
Case2-Liver-ROI3 &  65.94$\pm$27.01 &  64.20$\pm$26.53 &  1.74 & 0.66 & Case3-Stomach-ROI3 & 13.86$\pm$29.54 & 14.50$\pm$23.10 & -0.64 & 0.86 \\
Case2-Liver-ROI4 &  70.21$\pm$21.80 &  69.96$\pm$22.18 &  0.26 & 0.93 & Case3-Stomach-ROI4 & 25.40$\pm$32.31 & 24.40$\pm$24.75 & 1.01 & 0.77 \\
Case2-Liver-ROI5 &  70.56$\pm$26.19 &  68.02$\pm$21.97 &  2.55 & 0.44 & Case3-Stomach-ROI5 & 32.07$\pm$26.95 & 31.93$\pm$26.82 & 0.14 & 0.96 \\
\bottomrule
\end{tabular}}
\label{tab:2}
\footnotesize{$^\ddag$ROIs = Case number - measured organ - region of interest number.\\
$^\ast$ Data are mean $\pm$ standard deviations.\\
$^\dag$ $P<.05$ is defined as the significance level.\\}
\end{table}

\newpage
\section*{References}
\addcontentsline{toc}{section}{\numberline{}References}
\vspace*{-20mm}

% Following assumes you are using bibtex. However, for submission to the
% journal you MUST explicitly INCLUDE THE REFERENCES IN THE TEX FILE.
% In that case you need the following

% \begin{thebibliography}{10}
% insert the .bbl file generated by bibtex here
	%This will be a series of entries from your .bib file formatted
	%something like
	%\bibitem{Me09}
        %{I.~Meijsing, B.~W.~Raaymakers, A.~J.~E.~Raaijmakers \it et al.},
        %\newblock {Dosimetry for the MRI accelerator: the impact of a
	%magnetic field on the response of a Farmer NE2571 ionization chamber},
        %\newblock Phys. Med. Biol. {\bf 54}, 2993 -- 3002 (2009).

% \end{thebibliography}

% The following is when using bibtex and picks up the example.bib file

%\bibliography{Explicit address of .bib file}
%\bibliography{./pnas-sample}      %example.bib is on the same directory

% above points to where we find the master reference list
% and also causes the bibliography to be printed

% When creating your bibliography you should run bibtex on your local
% computer after running pdflatex on your .tex file. bibtex will
% generate a .bbl file.
% Copy the contents of this .bbl file into your main latex document,
% replacing the "\bibliography" command which was pointing at your .bib file.

% following defines style of .bbl file

%\bibliographystyle{explicit relative path to medphy.bst}
\bibliographystyle{./medphy}    %if this is installed on your system,
				    %it is not essential to have the    ./

% Note that you need to typeset once, then run bibtex, then typeset another
% two times to get the references working properly.

\end{document}